\documentclass[floatfix,pra,twocolumn,showpacs,amsmath,amssymb,letterpaper,groupaddresses,superscriptaddress]{revtex4-1}
\usepackage{times}
\usepackage{latexsym}
\usepackage{graphicx}
\usepackage{amsmath,amssymb}
\usepackage{verbatim,times,bbm}

\usepackage{color}

\newcommand{\be}{\begin{equation}}
\newcommand{\ee}{\end{equation}}
\newcommand{\bea}{\begin{eqnarray}}
\newcommand{\eea}{\end{eqnarray}}

\begin{document}

\title{Noise to lubricate qubit transfer in a spin network}

\author{Morteza Rafiee\footnote{m.rafiee178@gmail.com}}
\affiliation{Department of Physics, University of Shahrood, 3619995161, Shahrood, Iran}

\author{Cosmo Lupo\footnote{clupo@mit.edu}}
\affiliation{Massachusetts Institute of Technology, Research Laboratory of Electronics, 77 Massachusetts Avenue, Cambridge, Massachusetts 02139, USA}
\affiliation{School of Science and Technology, University of Camerino, I-62032 Camerino, Italy}

\author{Stefano Mancini\footnote{stefano.mancini@unicam.it}}
\affiliation{School of Science and Technology, University of Camerino, I-62032 Camerino, Italy}
\affiliation{INFN-Sezione di Perugia, Via A. Pascoli, I-06123 Perugia, Italy}

\begin{abstract}
We consider quantum state transfer in a fully connected spin network, in which the results indicate that it
is impossible to achieve high fidelity by free dynamics. However, the addition of certain kinds of noise can be
helpful for this purpose. In fact, we introduce a model of Gaussian white noise affecting the spin-spin couplings
(edges), except those linked to the input and output node, and prove that it enhances the fidelity of state transfer. The observed noise benefit is scale free as it applies to a quantum network of any size. The amount of the fidelity enhancement, depending on the noise strength as well as on the number of edges to which it is applied, can be so high as to take the fidelity close to one.
\end{abstract}


\maketitle

\section{Introduction}

A futuristic quantum version of the internet has recently attracted a lot of attention~\cite{qinternet}.
Experimental efforts have been devoted to realize proof-of-principle demonstrations of such networks
in various mesoscopic systems, such as photonic crystals~\cite{jeremy}, ion traps~\cite{nobel} and
superconducting circuits~\cite{franco}.
Multiuser quantum networks have the final goal of realizing a number of communication protocols.
The study of interacting qubit (spin-1/2) networks constitutes a good testing ground to this end.
These kind of networks have been considered as good candidates for engineering quantum channels
allowing the faithful transfer of quantum information between nodes~\cite{sourev}.
They turn out to be useful because they implement data buses by simply undergoing free dynamics after
an initial set-up. In this way, the possibility of having perfect state transfer (PST) comes from
suitable quantum interference effects in the coherent dynamics of the network.
However, PST requires the coupling constants to provide the right phase matching allowing the perfect
transfer of both amplitude and phase of a quantum state from one node to another. The conditions for
that to happen have been the subject of intense study, for a review see~\cite{AKay}.
It follows that if these conditions are not fulfilled dispersion effects and destructive interference
can make PST impossible. They determine the loss of information between communicating nodes and in some extreme
cases information can even remain localized, similarly to the Anderson localization effect~\cite{And}.
Some ways to circumvent this problem have been developed, e.g, based on local control of the couplings~\cite{BBBV}
or on protocols for lifting the encoding of qubits into multiparticle states~\cite{bb,os}.

Here we take a different approach, namely since too much coherence seems to be an obstacle to quantum communication, we shall exploit the addition of noise to reduce the interference effects, i.e. employ the latter as \emph{lubricant} for quantum information transmission \footnote{ Noise as 'quantum lubricant' has been originally discussed in the context of heat engines by T. Felman and R. Kosloff, Phys. Rev. E \textbf{73}, 025107(R) (2006).}.
This possibility has already been pointed out in \cite{FC10} following a series of papers aimed at showing that noise can foster excitation (i.e. classical information) transfer (see e.g. \cite{refdeph}).
There, onsite noise has been considered, that is noise affecting vertices of a quantum network. In contrast we shall consider noise affecting edges of a quantum network, i.e. intersite couplings.
Since disordered couplings were considered much more deleterious than disordered frequencies \cite{GDC05}, it is unforeseeable that such kind of noise can lead to benefits as well.

Actually we investigate the model of a fully connected qubit network, where each qubit interacts
with all the others with equal strength. First we show that in such a symmetric setup,
destructive interference does not allow PST across a given pair of nodes. Then we prove
that the addition of Gaussian white noise to spin-spin couplings (edges),
except those linked to the input and output node, is able to enhance
the fidelity of state transfer with respect to the free dynamics of the fully connected network.
It turns out that the noise benefit increases by increasing the strength of the noise, besides the number of edges to which it is applied.
In the strong-noise limit this effect can be explained as a consequence of the quantum Zeno effect~\cite{sav}.
Remarkably, an enhancement of the fidelity is observed even faraway from the Zeno limit,
for intermediate and small values of the noise strength.

\section{The network model}

Consider a simple undirected graph (that is, without loops or parallel edges) $G=(V,E)$,
with set of vertices $V(G)$ and set of edges $E(G)$.
Let $V(G)=\{1,\ldots,n\}$.
The adjacency matrix $A(G)$ is defined
by $[A(G)]_{kl}=1$, if $\{k,l\}\in E(G)$, and $[A(G)]_{kl}=0$ if $\{k,l\}\notin E(G)$.

Then a network of $n$ spin-$1/2$
quantum particles is realized by attaching particles to the vertices of $G$,
while representing their allowed couplings with the edges of $G$.
In this paper we consider the $XY$ interaction model so that $\{k,l\}\in E(G)$
means that the particles $k$ and
$l$ interact by the Hamiltonian $[H_G]_{kl}=\left(  \sigma^X_{k} \sigma^X_{l}
+\sigma^Y_{k} \sigma^Y_{l}\right)$.
Here $\sigma^X_{k}$, $\sigma^Y_{k}$ and $\sigma^Z_{k}$
denote the usual Pauli operators of the $k$-th particle, with $\sigma^Z_{k} |0\rangle_k = - |0\rangle_k$
and $\sigma^Z_{k} |1\rangle_k = |1\rangle_k$.
Moreover, we consider unit coupling constant. Then, the Hamiltonian of the whole network reads
\begin{equation}\label{Hnet}
H_G = \frac{1}{2}\sum_{k\neq l\in V(G)} [A(G)]_{kl}\left(  \sigma^X_{k}\sigma^X_{l}+\sigma^Y_{k}\sigma^Y_{l}\right)  ,
\end{equation}
and the relative Hilbert space is $\left(  \mathbb{C}^{2}\right)^{\otimes n}$.

The single excitation subspace is defined as the span of the vectors
$\{|1\rangle := |1\rangle_1 |0\rangle_2\cdots|0\rangle_n ,
|2\rangle := |0\rangle_1 |1\rangle_2\cdots|0\rangle_n , \dots,
|n\rangle := |0\rangle_1 |0\rangle_2\cdots|1\rangle_n\}$,
where the vector $|k\rangle$ indicates the presence
of the excitation on the $k$-th site and the absence on all the others.
In the basis $\{|k\rangle\}_{k=1,\dots,n}$, the
Hamiltonian~\eqref{Hnet} has entries $[H_G]_{kl}=2[A(G)]_{kl}$.
In the following we restrict our attention to the $n+1$ dimensional subspace generated
by the single excitation subspace together with the vector
$|0\rangle := |0\rangle_1 |0\rangle_2\cdots|0\rangle_n$,
belonging to the null space of $H_G$.

Let us take $\mathrm{i},\mathrm{o}$ such that $1 \le \mathrm{i} < \mathrm{o} \le n$ and indicate with them
the \emph{input} and \emph{output} nodes respectively.
We hence consider a generic qubit state for the $\mathrm{i}$-th node,
\be\label{psiini}
|\psi\rangle_\mathrm{i} = \cos\frac{\theta}{2} |0\rangle_\mathrm{i} +e^{i\phi}\sin\frac{\theta}{2} |1\rangle_\mathrm{i}
\ee
(where $\theta\in[0,\pi]$, $\phi\in[0,2\pi]$ and $i$ denotes the imaginary unit),
so that the initial state of the network reads
\be
|\psi(0)\rangle = |0\rangle_1 \dots |\psi\rangle_\mathrm{i}  \dots |0\rangle_n.
\ee
The time evolution of the density operator $\rho(t)$ with initial condition
\be
\rho(0)=|\psi(0)\rangle\langle\psi(0)|,
\label{rhoini}
\ee
can be described by the master equation
\be
\dot{\rho}(t)=\mathcal{L}\rho(t),
\label{me}
\ee
where $\mathcal{L}\rho=-i[H_G,\rho]$. Clearly,
for this kind of Liouvillian superoperator we can write
$\rho(t)=U_t \rho(0) U_t^{\dag}$ with $U_t=e^{-i H_Gt}$.
The state of the output qubit after time $t$ is described by the partial trace
overall qubits but the output one, $\rho_\mathrm{o}(t) = {\mathrm Tr}_{\not\mathrm{o}}[\rho(t)]$.
The resulting reduced dynamics corresponds to an amplitude damping channel
applied to the input state~\cite{sourev},
\be\label{Kraus_cg}
\rho_\mathrm{o}(t) = M_0 |\psi\rangle_\mathrm{i}\langle\psi| M_0^\dag + M_1 |\psi\rangle_\mathrm{i}\langle\psi| M_1^\dag,
\ee
with Kraus operators
\begin{eqnarray}
M_0 & = & |0\rangle_\mathrm{oi}\langle 0| + z |1\rangle_\mathrm{oi}\langle 1|,  \\
M_1 & = & \sqrt{1-|z|^2} \, |0\rangle_\mathrm{oi}\langle 1 |,
\end{eqnarray}
and
\begin{equation}
z := \langle \mathrm{o} | U_t | \mathrm{i} \rangle.
\end{equation}
The fact that the reduced dynamics is described by an amplitude damping channel
is a consequence of the commutativity of the system Hamiltonian~(\ref{Hnet}) with
the operator $N_G=\sum_{k=1}^n |k\rangle\langle k|$ expressing the total number
of excitations in the network.

PST is obtained if the state at the output node $\rho_\mathrm{o}(t)$ is a perfect reproduction of the input state
$|\psi\rangle_\mathrm{i}$.
In order to compare the states of qubits located at different nodes of the network
we have to map the Hilbert space of the input qubit into that of the output.
Such an identification is obtained by selecting a unitary transformation $V$ acting on
the $\mathrm{o}$-th qubit.
The \emph{input-output} fidelity~\cite{Richard} at time $t$ can be hence defined as
\be\label{fdef}
f_\mathrm{io}^V(\theta,\phi,t) := {}_\mathrm{i}\langle \psi | V \rho_\mathrm{o}(t) V^{\dag} | \psi \rangle_\mathrm{i},
\ee
and PST is obtained if there exist an evolution time $t$ and a unitary $V$
such that $f_\mathrm{io}^V(\theta,\phi,t) = 1$ for all $\theta$ and $\phi$.
As figures of merit we hence consider the average fidelity
\be\label{Fave}
F_\mathrm{io}^V(t) = \frac{1}{4\pi}\int_0^\pi d\theta \sin\theta \int_0^{2 \pi} d\phi \; f_\mathrm{io}^V(\theta,\phi,t),
\ee
and its maximum over $V$
\be\label{MFave}
F_\mathrm{io}(t) = \max_{V} F_\mathrm{io}^V(t).
\ee
In order to optimize over $V$, let us set
$V = u |0\rangle_\mathrm{io}\langle 0| + v |0\rangle_\mathrm{io}\langle 1| - v^* |1\rangle_\mathrm{io}\langle 0| + u^* |1\rangle_\mathrm{io}\langle 1|$,
with $u,v\in\mathbb{C}$, $|u|^2+|v|^2=1$, which  yields
\be
F_\mathrm{io}^V(t) = \frac{1}{2} + \frac{\mathfrak{Re}(z u^2)}{3} + \frac{|z|^2}{6}\left(2|u|^2-1\right).
\ee
It follows that the optimal choice is $u = e^{-i/2 \arg{z}}$, which accounts
for a local phase shift on the output qubit~\cite{sourev}, and gives
\be\label{Fsimple}
F_\mathrm{io}(t) = \frac{1}{2} + \frac{|z|}{3} + \frac{|z|^2}{6}.
\ee

\subsection{The Complete Graph}

A complete graph $K$ is such that every two vertices are adjacent. For this graph,
following Ref.~\cite{andrea2}, we can easily prove the impossibility of PST for $n>2$,
that is
\be
 \max_{t}F_\mathrm{io}(t) \, \left\{
 \begin{array}{cc}
 =1, & n=2,\\
 <1, & n>2.
 \end{array}\right.
 \label{exprop}
\ee
The Hamiltonian~\eqref{Hnet} associated to $K$ has the following components, in the basis
$\{|k\rangle\}_{k=1,\ldots,n}$ of the single excitation subspace,
\be
[H_K]_{kl}=
\left\{
\begin{array}{ccc}
0, & \mbox{if} \, \,  k=l,\\
2, & \mbox{if} \, \, k\neq l.
\end{array}
\right.
\label{adj}
\ee
Its eigenvalues are $\lambda_{1}=(2n-2)$ (with single degeneracy) and
$\lambda_{2}=-2$ (with degeneracy $n-1$).
The eigenvector with single degeneracy is $|\aleph\rangle = n^{-1/2} \sum_{k=1}^n |k\rangle$, hence
\begin{equation}
U_t = |0\rangle\langle 0| + e^{-i\lambda_{2}t} \sum_{k=1}^n |k\rangle\langle k| + \left( e^{-i\lambda_{1}t} - e^{-i\lambda_{2}t} \right)|\aleph\rangle\langle\aleph|.
\end{equation}
That allows us to evaluate
\begin{equation}\label{fid}
|z|^2= \frac{2}{n^2}\left[1-\cos(2nt)\right].
\end{equation}
Looking at~\eqref{Fsimple} we realize that the above quantity must
be equal to $1$ in order to have state transfer with unit fidelity.
Then Eq.~\eqref{exprop} immediately follows.

\section{Adding the noise}

Given that PST cannot be achieved in a complete graph $K$ one can devise
strategies to increase the fidelity~\eqref{MFave} eventually taking it close to 1.
Here we resort to the most counterintuitive mean, i.e., the addition of noise.
We take our cue from the fact
that in certain settings PST can
be achieved if the link between the $\mathrm{i}$-th and $\mathrm{o}$-th qubits is removed from
the complete graph~\cite{andrea1,andrea2}.
Then, besides the Hamiltonian~\eqref{Hnet} for a complete graph $K$,
which by virtue of adjacency matrix~\eqref{adj} can be written as
\begin{equation}\label{HK}
H_K = 2 \sum_{k < l\in V(K) } \left( |k\rangle \langle l| + |l\rangle \langle k| \right),
\end{equation}
we consider the addition of the following stochastic Hamiltonian
\begin{equation}\label{stochH}
{\mathfrak H}_m(t) = \sum_{k<l \in W_m(K)}  \xi_{kl}(t) \left( |k\rangle \langle l| + |l\rangle \langle k| \right),
\end{equation}
where $\xi_{kl}(t)$ are identical and independent Gaussian white noise terms, with
\begin{equation}\label{corr}
\langle \xi_{kl}(t)\rangle=0,\quad \langle \xi_{kl}(t)\xi_{k'l'}(t^{\prime})\rangle=2\eta\delta_{kk'}\delta_{ll'}\delta(t-t^{\prime}),
\end{equation}
and $\eta\geq 0$ measures the strength of such noise. Furthermore,
$W_m(K) \subset V(K)$ is a subset of $m\leq n-2$ vertices not containing $\mathrm{i}$ and $\mathrm{o}$.
That is, the Hamiltonian $\mathfrak{H}_m(t)$ randomly couples a subset of $m$ qubits but does not acts on the
input and output qubits. Hence it gives rise to $m(m-1)/2$ noisy edges not adjacent in $\mathrm{i}$ and $\mathrm{o}$ vertices.

Notice that due to the symmetry of $H_K$, the reduced dynamics on the $\mathrm{i}$-th and $\mathrm{o}$-th
qubits, generated by the total Hamiltonian $H_K + \mathfrak{H}_m(t)$, does not depend on the way the $m$ qubits are chosen.
Then, in the interaction picture the dynamics is governed by the equation
\begin{equation}\label{merhoI}
\dot\rho^I(t) = -i \left[ \mathfrak{H}_m^I(t) , \rho_I(t) \right].
\end{equation}
where
\begin{equation}\label{rhoint}
\rho^I(t) = e^{i H_K t} \, \rho(t) \, e^{-i H_K t},
\end{equation}
and
\begin{equation}
\mathfrak{H}_m^I(t) = e^{i H_K t} \, \mathfrak{H}_m(t) \, e^{-i H_K t}.
\end{equation}
Eq.~\eqref{merhoI} can be formally solved as
\begin{equation}
\rho^I(t)=\rho^I(0) - i \int_0^t \left[ \mathfrak{H}_m^I(t') , \rho^I(t') \right] dt'.
\end{equation}
Inserting such a solution back into~\eqref{merhoI} we get
\begin{equation}\label{medouble}
\dot\rho^I(t) = -i \left[ \mathfrak{H}_m^I(t) , \rho^I(0) \right] - \int_0^t \left[ \mathfrak{H}_m^I(t) , \left[ \mathfrak{H}_m^I(t') , \rho^I(t') \right] \right] dt'.
\end{equation}
By taking the average over noise realizations, accounting for~\eqref{corr}, and
returning back to the Schr\"{o}dinger picture, we get the master equation~\cite{GK1976}
\begin{equation}\label{newcalL}
{\mathcal L}(\rho) = -i \left[ H_K , \rho \right] - {\mathcal D}_m (\rho),
\end{equation}
with
\begin{equation}
{\mathcal D}_m(\rho) = -\eta \hspace{-3mm} \sum_{k < l \in W_m(K)} \left(L_{kl} \rho L_{kl}^{\dag}-\frac{1}{2}  L_{kl}^{\dag} L_{kl}\rho
-\rho\frac{1}{2}  L_{kl}^{\dag} L_{kl}\right),
\label{calD}
\end{equation}
and Lindblad operators
\begin{equation}\label{Lkl}
L_{kl}=|k\rangle\langle l|+|l\rangle\langle k|.
\end{equation}
It is worth noticing that the Hamiltonians~\eqref{HK} and~\eqref{stochH} commute with the total
number of excitations in the network, $N_G=\sum_{k=1}^n |k\rangle\langle k|$.
We could hence expect that the reduced dynamics can be expressed in a form analogous
to~(\ref{Kraus_cg}). However, the average over the noise realization induces dephasing
[as it is evident from the double commutator at the right-hand side~\eqref{medouble}],
and thus the reduced dynamics will be a combination of an amplitude damping and
a dephasing channel
\begin{eqnarray}
\rho_\mathrm{o}(t) &=&  M_{0+} |\psi\rangle_\mathrm{i}\langle\psi| M_{0+}^\dag + M_{0-} |\psi\rangle_\mathrm{i}\langle\psi| M_{0-}^\dag
\nonumber\\
&+& M_1 |\psi\rangle_\mathrm{i}\langle\psi| M_1^\dag ,
\label{damdeph}
\end{eqnarray}
where
\begin{eqnarray}
M_{0\pm} & = & \sqrt{\frac{1\pm\lambda}{2}} \, \left( |0\rangle_\mathrm{i}\langle 0| \pm z |1\rangle_\mathrm{i}\langle 1| \right),  \\
M_1    & = & \sqrt{1-|z|^2} \, |0\rangle_\mathrm{i}\langle 1 |,
\end{eqnarray}
and $\lambda \geq 0$.
For this kind of map we have
\be
F_\mathrm{io}^V(t) = \frac{1}{2} + \lambda \frac{\mathfrak{Re}(z u^2)}{3} + \frac{|z|^2}{6}\left(2|u|^2-1\right),
\ee
which, similarly to~(\ref{Fsimple}), yields
\begin{equation}\label{fid-ap}
F_\mathrm{io}(t) = \frac{1}{2} + \frac{\lambda|z|}{3} + \frac{|z|^2}{6}.
\end{equation}

\section{The resulting fidelity}

We are now going to relate $\lambda$, $z$, hence the fidelity $F_\mathrm{io}(t)$,
to the noise strength $\eta$ and quantify the effect of the noise.

\subsection{Four node network}

\begin{figure} \centering
    \includegraphics[width=8.5cm,height=5.5cm,angle=0]{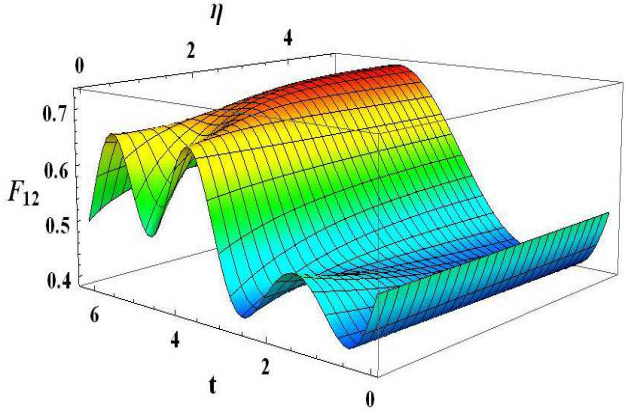}
    \caption{(Color online) Average fidelity $F_{{\mathrm io}}$ for state transfer
              between nodes 1 and 2 of the network with $n=4$ vs (dimensionless) time
              $t$ and (dimensionless) strength of the noise $\eta$ acting on the edge $3-4$.}
     \label{fig1}
\end{figure}

To have a clear picture of the effect of noise let us consider the
above model in the simplest configuration, i.e., a four-node fully
connected network.
Without loss of generality, we may consider nodes $1$ and $2$ as the $\mathrm{i}$
and $\mathrm{o}$ respectively and the noise is added to the $3$-$4$ edge.
In this case the dynamics can be explicitly solved, yielding the following
expressions for the parameters of the maximum average fidelity~\eqref{fid-ap}
\bea
|z|^2&=&2 e^{-\eta t /4}\left\{\left[\cosh\left(\frac{p t}{4}\right)+\frac{\eta}{p} \sinh\left(\frac{p t}{4}\right)\right]\right.
\nonumber \\
&&\left. -4 \cos(2 t)\left[\cosh\left(\frac{q t}{4}\right)+\frac{\eta}{q} \sinh\left(\frac{q t}{4}\right)\right]\right\}-\frac{3}{2},\nonumber\\
\label{z2}
\eea
and
\bea
\lambda z = \frac{e^{i t-\eta t/4}}{2} \left[\cosh\left(\frac{qt}{4}\right)+\frac{\eta}{q} \sinh\left(\frac{qt}{4}\right)\right]-\frac{e^{-i t}}{2},\nonumber \\
\label{lz}
\eea
with
\begin{eqnarray}
p & = & \sqrt{\eta^2-256}, \label{eqp} \\
q & = & \sqrt{\eta^2-64}. \label{eqq}
\end{eqnarray}
In Fig.~\ref{fig1} it is plotted~\eqref{fid-ap} using~\eqref{z2}, and ~\eqref{lz}.
It can be seen that at the time $3\pi/2$ the fidelity increases
with $\eta$ well beyond the maximum value it takes at $\eta=0$.
It is also worth noticing that the transition from purely imaginary $p,q$ to real $p,q$
changes trigonometric to hyperbolic functions in~\eqref{z2} and ~\eqref{lz}, making the behavior
of $F_\mathrm{io}(t)$ monotonically increasing vs $\eta$ up to reaching the unit value, which is a phenomenon
that will be discussed in the next section.

\subsection{Strong noise limit}

In this section we discuss the fidelity of the state transfer in the limit of strong
noise $\eta \gg 1$.
In the interaction picture the dynamics arising from~\eqref{newcalL} is described by the equation
\begin{equation}
\dot\rho^I(t) = \mathcal{D}^I_m(\rho^I(t)) \, ,
\end{equation}
where
\begin{equation}
{\mathcal D}^I_m(\rho^I(t)) = e^{i H_K t} \, {\mathcal D}_m(\rho^I(t)) \, e^{-i H_K t},
\label{calDint}
\end{equation}
and $\rho^I(t)$ is given by~\eqref{rhoint}.

One can easily realize that~\eqref{calD} leaves invariant the following subspace
${\mathsf S} = {\rm span}\left\{|0\rangle, |\mathrm{i}\rangle, |\mathrm{o}\rangle\right\}$.
In the strong noise limit the dynamics of the non-Hamiltonian term~(\ref{calD})
is much faster than the Hamiltonian one, hence we can look at the latter
as an adiabatic correction of the former.
As a consequence, if the network is initialized in a state belonging to the
invariant subspace ${\mathsf S}$ it will remain inside it
during all of the time evolution.
The initial state~\eqref{psiini} belongs to $\mathsf{S}$ and it will evolve inside
the instantaneous steady space
$\mathsf{S}_t = {\rm span}\left\{|0\rangle, e^{i H_K t} |\mathrm{i}\rangle,e^{i H_K t} |\mathrm{o}\rangle\right\}$,
that is,
\begin{equation}
\mathcal{D}^I_m \left( \rho_I(t) \right) = 0.
\end{equation}
Denoting by
\begin{equation}
P(t) := e^{i H_K t} \, P(0) \, e^{-i H_K t},
\end{equation}
the instantaneous projector onto $\mathsf{S}_t$, the adiabatic dynamics
is governed (in interaction picture) by the following equation~\cite{Kato}
\begin{equation}
\dot\rho^I(t) = - i [ \Xi(t) , \rho^I(t) ],
\end{equation}
where
\begin{equation}
\Xi(t) := i \left[ \dot P(t) , P(t) \right].
\end{equation}
By substituting
\begin{equation}
\dot P(t) = i \left[ H_K , P(t) \right],
\end{equation}
we obtain
\begin{equation}
\Xi(t) = 2 P(t) H_K P(t) - H_K P(t) - P(t) H_K.
\end{equation}
Finally coming back to the Schr\"odinger picture, we get
\begin{equation}
\dot\rho(t) = - i [ \widetilde{H}_K , \rho(t) ] ,
\end{equation}
where
\begin{align}
\widetilde{H}_K & := P(0) H_K P(0) + (1-P(0)) H_K (1-P(0)) \nonumber\\
& =  2 \sum_{k < l \not\in W_m(K)} \left( |k\rangle\langle l| + |k\rangle\langle l| \right). \label{Zenone}
\end{align}
Equation~(\ref{Zenone}) shows that the effect of a strong noise acting on the edges associated
to the subset $W_m(K)$ of $m$ vertices is to effectively {\it remove} the qubits belonging to $W_m(K)$.
That is, due to the Zeno effect the strong noise decouples the $m$ qubits from the rest of the network.
As a result, the fully connected network of size $n$ is mapped into a fully connected network of
size $n-m$. Notice that from~(\ref{Fsimple}) and~(\ref{fid}) one deduces that the state transfer fidelity in a fully
connected $XY$ network is a decreasing function of $n$. It hence follows that the Zeno effect, by
effectively reducing the size of the network, leads to an enhancement of the state transfer fidelity
\footnote{Actually, the Zeno effect leads to the enhancement
of the state transfer fidelity for the fully connected qubit network under more general conditions.
Indeed, any strong noise acting on the $m$ qubits will ultimately leads their effective removal from
the network, hence mapping the network of size $n$ to one of size $n-m$.}.
In the extreme case in which the noise acts on edges adjacent on all qubits but the input and output ones,
one reaches PST. In fact, in such a case, the initial state~\eqref{psiini}
will evolve into
\begin{equation}
|\psi(t)\rangle=\cos\frac{\theta}{2}|0\rangle+e^{i \phi} \sin\frac{\theta}{2}
\left( \cos t |\mathrm{i}\rangle+ i \sin t |\mathrm{o}\rangle\right),
\end{equation}
giving $z=i$ and $\lambda=1$. Notice that this result does not depend on either $m$ or $n$. Indeed, this is
a consequence of the fact that all pairs of adjacent nodes interact with the same
coupling constants in the Hamiltonian (\ref{Hnet}) and, in a fully connected network, each site is equivalent to another.

\subsection{Weak noise limit}

When the noise strength $\eta$ is much smaller than the
coupling parameter [assumed to be $1$ in~\eqref{HK}],
the density operator can be expanded in powers of $\eta$.
By truncating the expansion to the first order, we get
\begin{equation}\label{expansion}
\rho_{(2)}(t) = r_{(0)}(t) + \eta r_{(1)}(t).
\end{equation}
Substituting this expression into the master equation with
Liouvillian~\eqref{newcalL}, we obtain the equations:
\begin{align}
\dot{r}_{(0)}(t) = & -i [ H_K , r_{(0)}(t) ],\\
\dot{r}_{(1)}(t) = & -i [ H_K , r_{(1)}(t) ] - {\mathcal D}_m(r_{(0)}(t)).
\end{align}
Moving to the interaction picture [see~\eqref{rhoint} and~\eqref{calDint}], these equations become
\begin{align}
\dot r^I_{(0)}(t) & = 0,
\label{eqr0int} \\
\dot r^I_{(1)}(t) & = {\mathcal D}_m^I(r_{(0)}(0)).
\label{eqr1int}
\end{align}
They can be easily solved with initial state~\eqref{psiini}. Then
a straightforward calculation shows that the reduced density matrix of the
output qubit is as in Eq.~(\ref{damdeph}) with
\begin{eqnarray}
|z|^2 & = & |\beta|^2 + \eta \xi_1 \, , \\
\lambda z & = & |\beta|^2 +\eta \xi_2 \, ,
\end{eqnarray}
where
\begin{equation}
\beta = \frac{e^{it}}{n}(e^{-int}-1),
\label{eqb2}
\end{equation}
and
\begin{align}
\xi_1 &= m \Big\{ b_3 |\beta|^2 + b_4 \left[{\beta'}^* \beta + |\beta|^2m\right]
+ b_5 \left[\beta' \beta^* + |\beta|^2 m\right]\nonumber  \\
&\hspace{0.8cm} + b_6 [|{\beta'}|^2 + |\beta' \beta|^2 m^2 + |\beta|^2 m^2] \nonumber\\
&\hspace{0.8cm} + b_7 \left[{\beta'}^* + |\beta|^2\left(m^2+n-1\right)\right]\nonumber \\
&\hspace{0.8cm} + b_8 [|{\beta'}|^2 + |{\beta'} \beta|^2 \left(m^2+n-1\right) m^2(n-2)\nonumber \\
&\left. \hspace{1.6cm} + |\beta|^2 m(m^2+n-1)\right] \Big\}+ {\rm c.c.},
\label{eqx1}\\
\xi_2 & = m\left[b_1 \beta + b_2\beta'+b_2 \beta m\right],
\label{eqx2}
\end{align}

with
\begin{equation}
\beta' = \frac{e^{it}}{n}(e^{-int}+n-1),
\label{eqb1}
\end{equation}
and
\begin{eqnarray}
 b_1&=&(n-3)^2 \int_0^t d \tau \beta \beta'^*, \nonumber \\
 b_2&=&(n-3)^2 \int_0^t d \tau |\beta|^2, \nonumber \\
 b_3&=&(n-3)^2 \int_0^t d \tau \beta {\beta'}^* |{\beta'}|^2 , \nonumber \\
 b_4&=&(n-3)^2 \int_0^t d \tau (\beta^2 {\beta'}^{*2}  + |\beta'|^2 |\beta|^2 ),\nonumber \\
 b_5&=&(n-3)^2 \int_0^t d \tau |\beta|^2 |\beta'|^2 , \nonumber\\
 b_6&=&2(n-3)^2 \int_0^t d \tau \mathfrak{Re}{(\beta {\beta'}^*)} |\beta|^2,\nonumber \\
 b_7&=&(n-3)^2 \int_0^t d \tau \beta |\beta|^2 {\beta'}^*, \nonumber \\
 b_8&=&(n-3)^2 \int_0^t d \tau |\beta|^4.
\end{eqnarray}
The maximum average fidelity can then be computed using~\eqref{fid-ap}.
Let us introduce the following quantity
\begin{align}
\Delta(t, m,n, \eta) := \max\Big[ &F_\mathrm{io}(t; m, n,\eta)\nonumber\\
&-\max_t F_\mathrm{io}(t; m, n,0), 0\Big].
\end{align}
If it becomes strictly positive for some $t$, it is a signature of noise benefit also in the weak noise regime.
%%%%%%%%%%%%%%%%%%%%%%%%%%%%%%%%%%%%%%%%%%%%%%%%%%%%%%%%%%%%%%%%%%%%%%%%%%%%%
\begin{figure} \centering
     \includegraphics[width=8.5cm,height=7cm,angle=0]{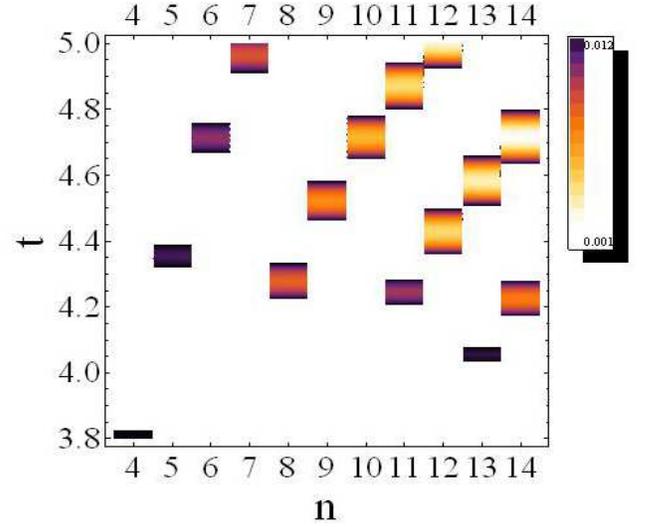}
    \caption{(Color online) Density plot of the quantity $\Delta$ vs $n$ and (dimensionless) $t$ for $\eta=0.01$ and $m=n-2$.
    Lighter regions correspond to smaller values of $\Delta$ (with blank corresponding to $0$).}
     \label{fig2}
\end{figure}
%%%%%%%%%%%%%%%%%%%%%%%%%%%%%%%%%%%%%%%%%%%%%
%%%%%%%%%%%%%%%%%%%%%%%%%%%%%%%%%%%%%%%%%%%%%%%%%%%%%%%%%%
In Fig.~\ref{fig2}, we show the density plot of  $\Delta$ vs $n$ and $t$ for $\eta=0.01$ and $m=n-2$, i.e.,
all edges affected by noise except those adjacent on the $\mathrm i$ and $\mathrm o$ nodes.
As can be seen, dark regions corresponding to greater than zero values can be found for any $n$
in a range of $t$ guaranteeing $\eta t\ll 1$.
Similar results can be found for larger values of $n$ as well.

\begin{figure} \centering
    \includegraphics[width=8cm,height=6.5cm,angle=0]{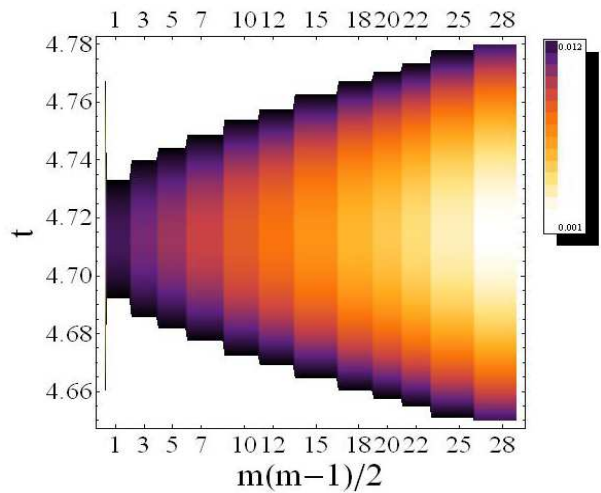}
    \caption{(Color online) Density plot of the quantity $\Delta$ vs $m(m-1)/2$ and (dimensionless) $t$ for $\eta=0.01$ and $n=10$.
    Lighter regions correspond to smaller values of $\Delta$ (with blank corresponding to $0$).}
     \label{fig3}
\end{figure}

Interestingly,  if we reduce the number of noisy edges, the benefit of noise persists up until $m=2$, i.e., one noisy edge, as can be seen in Fig.~\ref{fig3}.
Moreover, the magnitude of $\Delta$ does not significantly change with $m$. Only the width of the time window, over which $\Delta>0$ increases with $m$.

%%%%%%%%%%%%%%%%%%%%%%%%%%%%%%%%%%%%%%%%%%%%%%%%%%%%%%%%%%%%%%%%%%%%%%%%%%%%%

%%%%%%%%%%%%%%%%%%%%%%%%%%%%%%%%%%%%%%%%%%%%%%%%%%%%%%%%%%%%%%%%%%%%%%
%%%%%%%%%%%%

\section{Conclusion}

We have considered the problem of quantum state transfer in a fully connected qubit network.
Although PST is not allowed in the single excitation subspace under $XY$ dynamics, we have shown that the addition of noise facilitates the communication of quantum states over the network's nodes. In particular we have proved that the addition of Gaussian white noise to spin-spin couplings (edges), except those linked to the input and output node, enhances the fidelity of state transfer with respect to the free dynamics.
For strong noise, this can be seen as a consequence of the Zeno effect, that is, the introduction
of a strong noise acting on a certain subset of qubits decouples them to the rest of the network.
As a matter of fact, the effect of a strong noise acting on edges adjacent on $m$ qubits is to map the fully connected network of size $n$ into a smaller one of size $n-m$.
Since the state transfer fidelity of a fully connected $XY$ network is a decreasing function of the
network size, the presence of a strong noise term makes the transfer more favorable.
Remarkably, the benefit of the noise shows up as soon as the noise is introduced, although for small values of the noise strength the positive effect is tiny. Moreover, the advantage introduced by the noise can be considered as scale free in the sense that it persists independently of the network's size (although the amount of the effect may depend on $n$).
It is worth remarking that the strategy put forward is effective even if the noise terms are introduced
with different strengths $\eta_{kl}$ depending from the edge.
Vice versa, we suspect that colored noise would not be useful. This is because in case of  the non-Markovian dynamics there is usually a back flow of information to the system from the environment which tends to restore quantum interference. The latter is the main obstacle for qubit transfer in a highly connected network. However this aspect should be more deeply explored in future studies.

As for concern possible physical realizations we could mention a network of ions trapped in cavities connected by fibers. The addition noise on the links between the qubits can be artificially introduced into fibers by suitable rotators \cite{Alessandro}. Also a fully connected network could be implemented with superconducting qubits, as envisaged in \cite{Tsomokos}. Here the noise should be added on the harmonic oscillator circuit elements mediating the couplings.
Finally, we believe that the results found might also have implications in quantum biological complexes, where transport phenomena are strongly affected by noise \cite{harv}, and quantum gravity where the lattice structure of the spin system is not fixed \cite{qg}.

\acknowledgments

MR would like to thank the University of Camerino for hospitality.

%%%%%%%%%%%%%%%%%%%%%%%%%%%%%%%%%%%%%%%%%%%%%%%%%%%%%%%%

\end{document}